\documentclass[12pt]{article}
\usepackage{graphicx}

\begin{document}
\begin{titlepage}
%
%






{\ }

\begin{center}
\centerline{The chemical potential of a Lennard Jones fluid}

\end{center}
\begin{center}
\centerline{V.Celebonovic}			
\end{center}
\vskip5mm

\begin{center}
\centerline{Institute of Physics,Pregrevica 118,11080 Zemun-Belgrade,Serbia}
\vskip3mm 
\centerline{vladan@ipb.ac.rs}
\end{center}



\begin{abstract}
The aim of this paper is to present an analytical calculation of the chemical potential of a Lennard Jones fluid. The integration range is divided into two regions. In the small distance region,which is $r\leq\sigma$ in the usual notation,the integration range had to be cut off in order to avoid the occurence of divergences.In the large distance region,the calculation is technically simpler. The calculation reported here will be useful in all kinds of studies concerning phase equilibrium in a $LJ$ fluid. Interesting kinds of such systems are the giant planets and the icy satellites in various planetary systems,but also the (so far) hypothetical quark stars.\footnote{Published in Serbian Astron.Journal,{\bf181},pp.51-55,(2010)}
\end{abstract}
\end{titlepage}


{


\section{Introduction}

The aim of this paper is to present a calculation of the chemical potential of a fluid consisting of neutral atoms or molecules. Interest in such systems has considerably increased towards the end of the last century, as a consequence of progress in planetary science. Until the end of October 2010, according to data at http://exoplanet.eu, as much as $493$ planets outside the Solar system have been detected. It has been shown that $423$ of them have masses $M\leq5 M_{J}$ where $M_{J}$ is the mass of Jupiter. The major part ($398$) of the stars which have planets have masses equal to or smaller than the solar mass, and $151$ planet has semi-major axis of the orbit between $1$ and $3$ astronomical units. 

Judging by experience from our Solar System,it is expectable that this interval of distances from a star corresponds to temperatures under which fluids consisting of neutral atoms and molecules can exist. It is known that giant planets have huge atmospheres and dense fluid interiors. Another class of planetologically interesting systems,in which fluids are important,are the icy satellites in our planetary system. For example,it has been concluded from data accumulated in the course of the Galileo mission, that jovian satellites Europa and possibly Callisto  almost certainly have fluid oceans beneath their surfaces. 
    
Calculations to be discussed in this paper can find applications in theoretical studies of quark stars. These are (so far hypothetical) phases of extremely dense matter, expected to occur in the interiors of neutron stars. 
In a recent study,aiming to constrain the parameters of solid quark matter by using data on the binary pulsar $PSR J1614-2230$, the Lennard-Jones model was used to describe cold quark matter in quark stars [1] . It was shown there that if the number of quarks in one quark clusters is $N_{q}< 10^{3}$ there is enough parameter space for the existence of quark stars with masses higher than $2$ solar masses. 

Modelling theoretically the internal structure of celestial objects ranging from the icy satellites and/or the giant planets to quark stars, demands the knowledge of the chemical potential of the fluid which they contain in their interiors.

A necessary preparatory step in such a study must be the determination of the interparticle interaction potential. Obviously,for a fluid or any other kind of a system to be in equilibrium, the interparticle potential must be a combination of an attractive and a repulsive term. 

It is known that between a pair of neutral atoms or molecules at a mutual distance larger than their respective dimensions there exists an attractive force - called the van der Waals (vdW) force (for example [2] or [3] ). The potential corresponding to the vdW force is proportional to $r^{-6}$ ,where $r$ is the interparticle distance. As shown by F.London, the physical origin of the vdW forces is the interaction of instantenous multipoles,while the repulsive contribution is of electrostatic origin.

The vdW forces are anisotropic,which renders them additionally complicated [3]. However, their isotropic part is often approximated by the so called Lennard-Jones $LJ$ potential. All the calculations in the following will deal with this particular model potential.  
The $LJ$ model potential has the form
\begin{equation}
u(r)=4\epsilon\left[(\frac{\sigma}{r})^{12}-(\frac{\sigma}{r})^{6}\right] 
\end{equation} 
The symbol $\epsilon$ denotes the depth of the potential well,while $\sigma$ is the diameter of the molecular "hard core". Obviously, $lim_{r\rightarrow0}u(r)=\infty$. It can simply be shown that $lim_{r\rightarrow\sigma}u(r)= 0$ and that $(\partial u(r)/\partial r)=0$ for $r_{min}=2^{1/6}\sigma$. The depth of the potential well is $u(r_{min})= - \epsilon$. 

An example of the $LJ$ potential drawn  for the particular case of $CH_{4}$,with values of $\epsilon$ and $\sigma$ from [5], is represented on figure 1. This particular molecule is interesting in two research fields: planetary science, because it is present in the atmospheres of the giant planets, but also in studies of the interstellar medium. On the figure,the distance is expressed in units of $\sigma$ and the potential divided by the Boltzmann constant $k_{B}$ is given in the units of $K$. 
\begin{figure}
\centerline{\includegraphics[height=6.5cm]{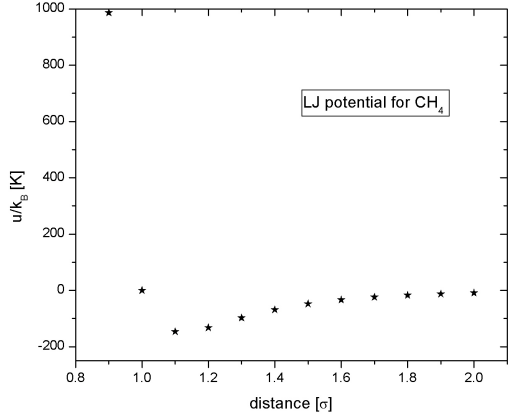}}
\begin{center}
\caption{The $LJ$ potential for methane ($CH_{4}$)}
\end{center}
\end{figure}

\section {The method of calculation of the chemical potential} 

The chemical potential of a fluid (or any other system) can be calculated in two different ways: by using the general thermodynamical formalism,or by the general formalism of statistical mechanics. 
\subsection{The thermodynamical formalism} 

In this approach the calculation starts from the definition of the Gibbs potential:
\begin{equation}
	G=U - TS + PV
\end{equation} 
where all the symbols have their standard meanings. Using the virial expansion, the pressure can be expressed as [4]: 
\begin{equation}
	P\cong P_{ID} (1+\frac{N}{V} B) 
\end{equation}
where $N$ is the number of particles in the system and $V$ the volume.The symbol $B$ denotes the second virial coefficient,given by:
\begin{equation}
	B=\frac{1}{2} \int_{0}^{\infty}(1-\exp^{- u(r)/T})dV 
\end{equation}
and $P_{ID}$ is the pressure of the ideal gas.The symbol $u(r)$ denotes the interaction potential. Inserting eq.(3) into eq.(2),it follows that
\begin{equation}
	G=U-TS+PV=G_{ID} +N P_{ID} B
\end{equation}

The chemical potential is defined as 
$\mu=(\partial G/\partial N)_{P,T}$,which implies that 
\begin{equation}
	\mu = (\frac{\partial G_{ID}}{\partial N})_{P,T} + P_{ID} B + N B (\frac{\partial P_{ID}}{\partial N})_{P,T}
\end{equation}
or 
\begin{equation}
	\mu=\mu_{ID} +P_{ID}B + N B (\frac{\partial P_{ID}}{\partial N})_{P,T}
\end{equation}
The equation of state of the ideal gas is $P_{ID}V = N T$ which finally leads to
\begin{equation}
	\mu = \mu_{ID} + 2 P_{ID} B
\end{equation}
For the particular case of the $LJ$ potential, it can be shown that the second virial coefficient is given by [5]: 
\begin{equation}
B(T)=-(b_{0}/2) \sum_{n=0}^{\infty}\frac{1}{ n!}\Gamma(\frac{2n-1}{4})(\frac{\epsilon}{T})^{\frac{2n+1}{4}}
\end{equation}
where $b_{0}=2\pi\sigma^{3}/3$ and $\Gamma$ denotes the Gamma function. Inserting eq.(9) into eq(8) it follows that
\begin{equation}
	\mu=\mu_{ID}-b_{0}p_{ID}\sum_{n=0}^{\infty}\frac{1}{ n!}\Gamma(\frac{2n-1}{4})(\frac{\epsilon}{T})^{\frac{2n+1}{4}} 
\end{equation}
which is the result for the chemical potential. Limitng the sum in this expression to terms up to and including $n=1$,it follows that:
\begin{eqnarray}
\mu \cong \mu_{ID}+2p_{ID}b_{0}\left[2.45083-1.8128(\frac{\epsilon}{T})^{1/2}\right]\nonumber\\
\times(\frac{\epsilon}{T})^{1/4}
\end{eqnarray}

\subsection{The formalism of statistical mechanics} 

The chemical potential of a fluid is given by [6] :
\begin{eqnarray}
	\frac{\mu}{k_{B}T}=\ln(\rho\lambda^{3})+\frac{\rho}{k_{B}T} \int_{0}^{1}d\gamma\int_{0}^{\infty}dr \nonumber\\
\times 4\pi r^{2} u(r) g(r)
\end{eqnarray}
where $\gamma$ denotes the "coupling parameter" [6],$\rho$ is the particle number density ,$u(r)$ the interaction potential and $g(r)$ is the radial distribution function. The symbol $\hbar$ is the Planck constant divided by $2\pi$,$m$ is the particle mass and $\lambda$ is the thermal wavelength given by
	\[\lambda=(\frac{2\pi\hbar^{2}}{m k_{B} T})^{1/2}
\]
Expression (10) is valid under the condition $\rho\lambda^{3}>1$,which leads to: 
\begin{equation}
	\rho\geq\left(\frac{m k_{B} T}{2\pi\hbar^{2}}\right)^{3/2}
\end{equation}

The radial distribution function is a "bridge" relating macroscopic thermodynamic properties with interparticle interactions in any kind of a substance. In the theory of liquids, $g(r)$ can be determined from first principles [6] just assuming a suitable form of the intermolecular potential [7]. In the following, the result for $g(r)$ obtained in [7] will be used.
Changing the variable from $r$ to $x=r/\sigma$,and performing the integration over $\gamma$,it follows that 
\begin{equation}
\frac{\mu}{k_{B}T}=\ln(\rho \lambda^{3})+4\pi\sigma^3 \frac{\rho}{k_{B}T} \int_{0}^{\infty}dx x^{2} u(x) g(x)	
\end{equation}
The domain of integration can be divided into two subdomains: $x\in[0,1]$ and $x\in[1,\infty]$,which means:
\begin{eqnarray}
	I=\sigma^{3}\int_{0}^{\infty}dx x^{2} u(x) g(x)=\nonumber\\
\sigma^{3}[\int_{0}^{1}dx x^{2} u(x) g_{1}(x)+\int_{1}^{\infty}dx x^{2} u(x) \times g_{2}(x)]\nonumber\\
  = \sigma^{3}\times[I_{1}+I_{2}]
\end{eqnarray}
This divergence of the LJ potential which occurs when $x\rightarrow0$ can be bypassed either by introducing a suitable change of the range of integration $x\in[x_{0},1]$ instead of $x\in[0,1]$ with $x_{0}\neq0$,or by changing the form of the potential in the domain $x\in[0,1]$.
For $x\in[0,1]$ the function $g(r)$ has the form 
\begin{equation}
	g_{1}(x)=s\exp[-(mx+n)^{4}] 
\end{equation}
and  for $x\in[1,\infty]$ the radial distribution function is 
\begin{eqnarray}
	g_{2}(x)=1+\frac{1}{x^{2}}\exp[-(ax+b)]\sin[(cx+d)]+\nonumber\\
\frac{1}{x^{2}}\exp[-(gx+h)]\cos[(kx+l)]
\end{eqnarray}
where $a$,$b$,$c$,$d$,$g$,$h$,$k$,$l$,$m$,$n$ and $s$ are functions of pressure, temperature and density given in [7]. 

The appropriate boundary conditions,namely that the radial distribution function should tend to 1 in the limits of zero density and infinite distance, and the consequences of these conditions are also discussed there. As a consequence,the functions $b$, $d$, $h$ and $l$ are functions of density only, $n$ is the function of temperature only and the other functions depend on the temperature and density [7].    

\section{The calculation}
\subsection{The case $x\in[0,1]$}

With the change of variables $x=r/\sigma$,the $LJ$ potential gets the form
\begin{equation}
	u(x)=4\epsilon[x^{- 12}-x^{-6}]
\end{equation}

Inserting Eqs (16) and (18) into the expression for $I_{1}$ in eq.(15),it follows that

\begin{eqnarray}
I_{1}=4 s \epsilon \sum_{l=0}^{\infty}\frac{(-1^{l})}{(l!)}\int_{x=x_{0}}^{1}(\frac{1}{x^{10}}-\frac{1}{x^4})\nonumber\\
(m x+n)^{4l} dx
\end{eqnarray}

Performing the integrations,after some algebra, it finally follows that
\begin{eqnarray}
I_{1}\cong\frac{8}{9}s\epsilon\times[1+\frac{27}{5}m^{4}-\nonumber\\
\frac{27}{10}m^{8}-\frac{9m^{12}}{54}+..+\frac{1}{2x_{0}^{9}}-\frac{n^{4}}{2x_{0}^{9}}+..]
\end{eqnarray}

\subsection{The case $x\in[1,\infty]$}

In this case,the calculation of the chemical potential is more straightforward.
Inserting Eqs.(17) and (18) into the expression for $I_{2}$ in eq.(15), and performing the integration,gives the following approximate result for the integral $I_{2}$: 
\begin{eqnarray}
I_{2}\cong -\frac{8\epsilon}{9}+\pi\epsilon\cos[d]\cosh[b][\frac{a^{5}}{120}-\frac{a^{3}c^{2}}{12}\nonumber\\+\frac{ac^{4}}{24}-\frac{c}{5}\cos[c]\cos[d]+\ldots]
\end{eqnarray}
\subsection{The chemical potential}

According to Eqs.(14) and (15) the chemical potential is given by
\begin{equation}
\mu=k_{B}T \ln(\rho\lambda^{3})+4\pi\rho\sigma^{3}(I_{1}+I_{2})
\end{equation}
where the first terms of $I_{1}$ and $I_{2}$ are given by Eqs.(20) and (21). 

Inserting Eqs.(20) and (21) into Eq.(22),one gets a simple analytical approximation for the chemical potential of a LJ fluid.

\section{Discussion and conclusions}

In this paper we have obtained an approximate analytical expression for the chemical potential of a Lennard Jones fluid.Two ways in which such an expression can be obtained have been presented,and both of these approaches has been applied. 

The approach based on the general thermodynamic formalism gives a result,expressed as eq.(10), which is both mathematically and physically simpler. It contains just two variables which characterize the  material under consideration - these are $\sigma$ - the diameter of the molecular "hard core", and $\epsilon$ - the depth of the potential well. Note that the chemical potential obtained in this way for a certain value of the ratio $\epsilon/T$ reduces to the value $\mu_{ID}$. 

The formalism of statistical mechanics is both mathematically and physically more complex.
The general conclusion is that the chemical potential depends on the thermodynamic parameters of the fluid through the functions $a$-$s$, which are in turn functions of the pressure and/or density and/or temperature [7] ,but also on the interaction parameters. The approximate expression for the chemical potential of a LJ fluid is:
\begin{eqnarray}
\mu \cong k_{B}T\ln(\rho\lambda^{3})+\frac{32}{9}\pi\rho\epsilon\sigma^{3}s[1+\frac{27m^{4}}{5}\nonumber\\+..+\frac{1}{2x_{0}^{9}}+..+\frac{a^{3}}{s}\cos[d]\cosh[b](\frac{3a^{2}}{320}-\frac{3c^{2}}{32})\nonumber\\
+..]
\end{eqnarray}
All the symbols in this expression have their standard meanings,or were introduced in (Morsali et al.,2005). Mathematically, the symbol $x_{0}$ denotes the cut off radius of the LJ potential introduced in the calculations in order to avoid the occurence of divergences. Physically,this quantity represents the interparticle distance at which pressure ionisation occurs. Qualitatively speaking, pressure excitation and/or ionisation occur because electronic energies change under the influence of the external pressure field. For details about this process see,for example,[8].   

The calculation presented in this paper was motivated by recent advances in planetary science. As a consequence of numerous discoveries of giant exoplanets, modellisation of their internal structure has regained importance. These planets consist mostly of fluids,and accordingly an obvious need for theoretically "preparing the ground" for the modellisation of their interiors has occured. Studies of phase equilibrium and phase transitions demand an explicit knowledge of the chemical potential. Some preliminary results in that direction have recently been obtained [9] in the limit of small density and without taking into account the chemical potential. Another interesting problem,which becomes accessible for study with the results obtained in this paper is the behaviour of the chemical potential of a LJ fluid with changes of its thermodynamical parameters. Some aspects of both of these problems will be discussed in future work.  


\section{Acknowledgement} 

The preparation of this work was financed by the Ministry of Science and Technology of Serbia under its project 141007. 

{}
\end{document}